\documentclass[a4paper,reqno]{amsart}
\usepackage{amsthm}
\usepackage{amsmath}
\usepackage{amssymb}
\usepackage{tikz}
\usepackage{color}
\usetikzlibrary{matrix}
\usepackage{dsfont}
\usepackage{graphicx}
\usepackage[margin=1.2 in]{geometry}
\usepackage{stackrel}
\usepackage{enumerate}
\usepackage{dashrule}
\usepackage{arydshln}
\usepackage{doi}
\numberwithin{equation}{section}

\theoremstyle{plain}
\newtheorem{thm}{Theorem}[section]

\newtheorem{lem}[thm]{Lemma}

\theoremstyle{definition}
\newtheorem{df}[thm]{Definition}

\theoremstyle{remark}
\newtheorem{rem}[thm]{Remark}

\theoremstyle{question}
\newtheorem{qst}[thm]{Question}

\newcommand{\bN}{{\mathbb N}}

\newcommand{\bR}{{\mathbb R}}

\newcommand{\bC}{{\mathbb C}}

\newcommand{\Tr}{\mathrm{Tr}}

\newcommand{\ba}{\begin{array}}
\newcommand{\ea}{\end{array}}
\newcommand{\be}{\begin{eqnarray*}}
\newcommand{\ee}{\end{eqnarray*}}
\newcommand{\beg}{\begin{eqnarray}}
\newcommand{\eeg}{\end{eqnarray}}
\newcommand{\beq}{\begin{equation}}
\newcommand{\eeq}{\end{equation}}
\newcommand{\beqn}{\begin{equation*}}
\newcommand{\eeqn}{\end{equation*}}

\newcommand*\oline[1]{%
  \vbox{%
    \hrule height 0.5pt
    \kern0.25ex
    \hbox{%
      \kern-0.1em
      \ifmmode#1\else\ensuremath{#1}\fi
      \kern-0.1em
    }
  }
}

\begin{document}

\title{A factorization property of positive maps on $C^*$-algebras.}

\begin{abstract}

 The purpose of this short note is to clarify and present a general version
of an interesting observation by  Piani and Mora (Physic.  Rev. A 75, 012305 (2007)) \cite{PM}, linking complete positivity of linear maps on matrix algebras to decomposability of their ampliations.

Let $A_i$, $C_i$ be unital C*-algebras and let $\alpha_i$ be
positive  linear maps from $A_i$ to $C_i,$ $i=1,2$.  We obtain
conditions under which any positive  map $\beta$ from the minimal
C*-tensor product $A_1 \otimes_{min} A_2$ to $C_1 \otimes_{min}
C_2$, such that  $ \alpha_1 \otimes \alpha_2 \geq \beta$, factorizes
as $\beta = \gamma \otimes \alpha_2$ for some positive map $\gamma$.
In particular we show that when
 $\alpha_i \colon A_i \rightarrow B(\mathcal H _i)$ are completely positive (CP) maps for some Hilbert
  spaces $\mathcal H _i$ $(i=1,2)$, and
  $\alpha_2$
  is a pure CP map and $\beta$ is a CP map so that $\alpha_1 \otimes \alpha_2 -
  \beta$ is also CP,
   then $\beta = \gamma \otimes \alpha_2$ for some CP map $\gamma$. We  show that a similar result holds in
   the context of positive linear maps when  $A_2 = C_2 = B(\mathcal H)$ and $\alpha_2 = id$.  As an application
we extend \cite[IX Theorem]{PM}( revisited recently by Huber et al in \cite{HLLM})  to show that for any
linear map $\tau$ from a unital C*-algebra $A$ to
 a C*-algebra $C$, if  $\tau \otimes id_k$ is decomposable for some $k \geq 2$, where $id_k$ is the identity map on
 the algebra $M_k(\bC )$ of $k\times k$ matrices,  then $\tau $ is completely positive.
\end{abstract}

\author{B. V. Rajarma Bhat}

\address{Indian Statistical Institute, R V College Post, Bangalore-560059, India}
\email{bhat@isibang.ac.in, bvrajaramabhat@gmail.com}

\author{Hiroyuki Osaka}


\address{Department of Mathematical Sciences\\
Ritsumeikan University\\
Kusatsu, Shiga 525-8577, Japan}
\email{osaka@se.ritsumei.ac.jp}


\date{\today}
\thanks{The first author's research was partially supported by J C
Bose Fellowship. The second author's research was partially
supported by KAKENHI Grant Number JP17K05285. This research was done
during the visit of the first author to  Ritsumeikan and Kyoto
Universities and the author gratefully acknowledges the hospitality
of his hosts H. Osaka and B. Collins.}

\maketitle

\section{Introduction}

It is now well established that completely positivity is an essential property for quantum channels and linear maps which are positive but not completely positive play an important role in studying entanglement as entanglement witnesses. These are primary justifications for studying such
linear maps in the context of quantum information theory. We begin with some basic definitions.

       Let $A$ and $C$ be C*-algebras and let $\phi\colon A \rightarrow C$ be a
linear map. For $k \in \bN,$ $\phi$ is said to be $k$-positive if
the map $\phi \otimes id_k\colon A \otimes M_k(\bC) \rightarrow C
\otimes M_k(\bC)$ defined by $(\phi \otimes id_k)([a_{ij}]) =
[\phi(a_{ij})]$ is positive, where $id_k$ denotes the identity map
on $M_k(\bC)$. A linear map $\phi$ is said to be completely positive
(CP) if $\phi$ is $k$-positive for all $k \in \bN$. Similarly,
$\phi$ is said to be $k$-copositive if the map $\phi \otimes t
\colon A \otimes M_k(\bC) \rightarrow C \otimes M_k(\bC)$ defined by
$(\phi \otimes t)([a_{ij}]) = [\phi(a_{ji})]$ is positive, where $t$
denotes the transpose map on $M_k(\bC)$. A linear map $\phi$ is said
to be completely copositive if $\phi$ is $k$-copositive for all $k
\in \bN$. A positive map is called decomposable if it is the sum of
a completely positive map and a completely copositive map. The first
example of an indecomposable positive linear map in $M_3(\bC)$ was
found by Choi \cite{Choi 1975}. Since then many other examples of
indecomposable positive linear maps have been found. But very little
is known about the structure of general positive linear maps. We
know that any positive linear map from $M_3(\bC)$ to $M_3(\bC)$ is a
decomposable map or an atomic map, meaning a map which can not be
written as a sum of a $2$-positive map and a $2$-copositive map by
\cite{YLT 2018}. In \cite{Terhal 2001} Terhal posed a question about
existence of  indecomposable $k$-positive maps  from $M_m(\bC)$ to
$M_n(\bC)$ for $1 < k < m  \leq  n$. Note that any 2-positive map
from $M_3(\bC)$ to $M_3(\bC)$ is decomposable \cite{YLT 2018}. Piani
and Mora \cite[IX Theorem]{PM} proved that a $k$-positive linear map
$\phi : M_m(\bC)\to M_n(\bC )$, is completely positive if and only
if $\phi \otimes id_k$  is decomposable. This gives us a convenient
tool to construct some positive maps which are not decomposable.
Recently
  M. Huber, L. Lami, C. Lancien and A. Muller \cite{HLLM}
have found this useful in their work.  The current article
came out of our efforts to understand this result.

   In this note  at first we consider a factorization
property for positive maps on C*-algebras and show that for any
linear map from a C*-algebra $A$ to a C*-algebra $C$ such that
$\alpha \otimes id$ majorizes a positive linear map $\beta\colon A
\otimes B(\mathcal H) \rightarrow C \otimes B(\mathcal H)$ there is a positive linear
map $\gamma\colon A \rightarrow C$ such that $\beta = \gamma \otimes
id$. As an application, we generalize the result of Pian and Mora to arbitrary
$C^*$-algebras. This also throws up some open questions.

\section{Main result}

We consider complex Hilbert spaces $(\mathcal H, \langle \cdot ,
\cdot \rangle )$, where the inner product is  anti-linear in the
first variable. For $x, y$ in $\mathcal H$, $|x\rangle \langle y|$
denotes the linear map on $\mathcal H$  defined by
$$|x\rangle \langle y|(z)= x\langle y, z\rangle .$$

It is well-known that the rank one projections are extremal in the
convex cone of positive operators of a Hilbert space. This
observation leads to the following simple result and we omit its
proof.

\begin{lem}\label{lem:majorization}
Let $C$ be a unital C*-algebra and let $\mathcal H$ be a Hilbert space. If $y \in C$, $h \in \mathcal H$ and $z \in C \otimes B(\mathcal H )$
satisfy $0 \leq z \leq y \otimes |h\rangle \langle h|$, then $z = v \otimes |h\rangle \langle h|$ for some $v \in C$.
\end{lem}

\vskip 2mm

Now we are ready to the prove the following factorization property
for positive maps.
\begin{thm}\label{factorization1}
Let $A$, $C$ be C*-algebras and let $\mathcal H$ be a Hilbert space.
Suppose that $\alpha\colon A \rightarrow C$ is a linear map and
$\beta:A \otimes B(\mathcal H) \rightarrow C \otimes B(\mathcal H)$
is a positive linear map. Suppose that $\alpha \otimes id - \beta$
is positive. Then $\beta = \gamma \otimes id$ for some positive
linear map $\gamma\colon A \rightarrow C$.
\end{thm}

\begin{proof}
For any $a \in A$ with $a\geq 0$ and $h \in \mathcal H$, we have
$$
0 \leq \beta(a \otimes |h\rangle \langle h|) \leq \alpha(a) \otimes |h\rangle \langle h|.
$$
Making use of
Lemma \ref{lem:majorization}, $\beta(a \otimes
|h\rangle\langle h|) = \gamma_h(a) \otimes |h\rangle \langle h|$ for
some $\gamma_h(a) \in C$. We need show that $\gamma_h(a)$ is
independent of $h$. This follows from the linearity of $\beta$.
Indeed, if $g, h \in \mathcal H$ and $g\perp h$, we get
\begin{align*}
\beta(a \otimes |g\rangle \langle g|) &= \gamma_g(a) \otimes |g\rangle \langle g|,\\
\beta(a \otimes |h\rangle \langle h|) &= \gamma_h(a) \otimes |h\rangle \langle h|,\\
\beta(a \otimes |g+h\rangle \langle g+h|) &= \gamma_{g+h}(a) \otimes |g+h\rangle \langle g+h|,\\
\beta(a \otimes |g-h\rangle \langle g-h|) &= \gamma_{g-h}(a) \otimes |g-h\rangle \langle g-h|.
\end{align*}
Adding last two equalities, we have
\begin{align*}
2\beta(a \otimes |g\rangle \langle g| ) + 2\beta(a \otimes |h\rangle \langle h|)
&= (\gamma_{g+h}(a) + \gamma_{g-h}(a))\otimes (|g\rangle\langle g| + |h\rangle \langle h|)\\
&+ (\gamma_{g+h}(a) - \gamma_{g-h}(a))\otimes (|g\rangle\langle h| +
|h\rangle \langle g|).
\end{align*}
So,
\begin{align*}
2\gamma_g(a) \otimes |g\rangle \langle g|  + 2\gamma_h(a) \otimes |h\rangle \langle h|
&= (\gamma_{g+h}(a) + \gamma_{g-h}(a))\otimes (|g\rangle\langle g| + |h\rangle \langle h|)\\
&+ (\gamma_{g+h}(a) - \gamma_{g-h}(a))\otimes (|g\rangle\langle h| +
|h\rangle \langle g|).\end{align*}
Comparing both sides, we get
\begin{align*}
&\gamma_g(a) = \frac{\gamma_{g+h}(a) + \gamma_{g-h}(a)}{2} = \gamma_h(a),\\
&\gamma_{g+h}(a) = \gamma_{g-h}(a).
\end{align*}
Therefore, for any $a \in A$ and $g \in \mathcal H$ we have $\beta(a
\otimes |g\rangle \langle g|) = \gamma(a) \otimes |g\rangle \langle
g|$ for some $\gamma (a).$  The linearity of $\gamma $ comes from
the linearity of $\beta$. Since $\beta(a \otimes |g\rangle \langle
g|)$ is positive, $\gamma(a) \otimes |g\rangle \langle g|$ is
positive, hence $(id_C \otimes \Tr)(\gamma(a) \otimes |g\rangle
\langle g|)$ is positive, and $\gamma(a)$ is positive.
\end{proof}

\vskip 2mm








It may be noted that, this Theorem and its proof can  easily be
modified to have a similar result when the identity map in the
second factor is replaced by a map of the form $x\mapsto mxm^*$ for
any $m\in B(\mathcal H).$ We wish to extend this idea further to a
more general setting and to this end we make the following
Definition.

\begin{df}
A CP map $\alpha$  {\em dominates\/} another CP map $\gamma$ if
$\alpha - \gamma$ is also CP. A CP map $\alpha\colon A \rightarrow
B(\mathcal H)$ is {\em pure \/} if
 whenever $\alpha$ dominates $\gamma$, then $\gamma = c \alpha$ for $c \in \bR_+$.
\end{df}

As observed by Arveson \cite{Arveson 1969}, the structure of
dominated maps and purity can be described in terms of the
Stinespring dilation  \cite{Stine} of the completely positive map as follows.
Suppose $(\mathcal K, \pi , V)$ is a minimal Stinespring triple of
the given CP map $\alpha : A\to B(\mathcal H)$, that is, $\mathcal
K$ is a Hilbert space, $\pi : A\to B(\mathcal K )$ is a
$*$-homomorphism and $V:\mathcal H\to \mathcal K$ is a linear map
such that (i) $\alpha (\cdot )= V^*\pi (\cdot )V$; and (ii)
$\mathcal K =~~\overline{\mbox{span}}~~\{ \pi (a)Vh: a\in A, h\in
\mathcal H\}.$ Then $\alpha $ dominates a completely positive map
$\gamma : A\to B(\mathcal H)$ if and only if $\gamma (\cdot )=
V^*\pi (\cdot )zV$ for some $z\in B(\mathcal K)$ satisfying $0\leq
z\leq I$ and $z\in \pi (A)'.$ Such a $z$ is unique. Further $\alpha$
is pure if and only if $(\pi(A))^{'} = \bC I_{\mathcal K}$ by
\cite[Corollary 1.4.3]{Arveson 1969}.

\vskip 2mm

\begin{thm}\label{factorization2}
Let $A_i$ be unital C*-algebras and let $\mathcal H_i$ be Hilbert
spaces $(i = 1,2)$. Let $\alpha_i\colon A_i\rightarrow B(\mathcal H
_i)$ $(i= 1, 2)$ be completely positive maps and $\alpha_2$ be pure.
Suppose that $\beta\colon A_1 \otimes_{\min}A_2 \rightarrow B(
\mathcal H _1) \otimes_{min} B(\mathcal H _2)$ is completely
positive and $(\alpha_1 \otimes \alpha_2)$ dominates $\beta$. Then,
$\beta = \beta_1 \otimes \alpha_2$ for some completely positive map
$\beta _1$, which is dominated by $\alpha_1$.
\end{thm}

\begin{proof}
Let $(\mathcal K _i, \pi_i, V_i)$ be minimal Stinespring triples of
$\alpha_i$ $(i=1, 2)$ respectively. Then for each $a_i \in A_i$ $(i
= 1, 2)$

$$
(\alpha_1 \otimes \alpha_2)(a_1 \otimes a_2) = (V_1 \otimes V_2)^*(\pi_1(a_1) \otimes \pi_2(a_2)(V_1 \otimes V_2).
$$
As $\beta$ is dominated by $\alpha_1 \otimes \alpha_2$, there exists $z \in (\pi_1(A_1)
 \otimes_{\min}\pi_2(A_2))^{`} \cong (\pi_1(A_1))^{'} \overline{\otimes_{\min}} (\pi_2(A_2))^{'}
\cong  (\pi_1(A_1))^{'} \otimes_{\min}\bC I_{\mathcal K _2}$ by
\cite[Proposition 4.13]{Takesaki 1979}, Tomita's Theorem, and the
pureness of $\alpha_2$, where $\overline{\otimes_{\min}}$ means the
von Neumann tensor product. Hence, there is $z = z_1 \otimes
I_{\mathcal K _2}$ for some $z_1 \in (\pi_1(A_1))^{'}$ and
\begin{align*}
\beta(a_1 \otimes a_2) &=  (V_1 \otimes V_2)^*(\pi_1(a_1) \otimes \pi_2(a_2)(z_1 \otimes
 I_{\mathcal K _2})(V_1 \otimes V_2)\\
&= \beta_1(a_1) \otimes \alpha_2(a_2),
\end{align*}
where $\beta_1(a_1) = V_1^*\pi_1(a_1)z_1V_1$. Therefore, we have
$\beta = \beta_1 \otimes \alpha_2$ such that $\beta_1$ is dominated
by $\alpha_1$.
\end{proof}

We are left with the following question where the domination and
purity should be in the class of positive linear maps. One has to be
careful in tackling this question as tensor products of positive
linear maps in general need not be positive.
\begin{qst}\label{qst:split}
Let $A_i, C_i, i=1,2$ be unital $C^*$-algebras and let $\alpha _i:A_i\to
C_i$ be positive linear maps and $\alpha _2$ is pure. Suppose $\beta
:  A_1 \otimes_{\min} A_2 \rightarrow C_1\otimes_{\min} C_2$ is a
positive linear map dominated by $\alpha _1\otimes \alpha _2.$ Does
it follow that there exists a positive linear map $\beta _1: A_1\to
C_1$ such that $\beta = \beta _1\otimes \alpha _2$?
\end{qst}

\section{Application}

As an application of ideas in the previous section we give a
different proof of an interesting  result by Piani and Mora \cite[IX
Theorem]{PM} and by Huber et al. \cite{HLLM}. Moreover the result is
generalized to arbitrary $C^*$-algebras.

\begin{thm}\label{thm:main2}
Let $A$ and $C$ be unital C*-algebras and $\tau\colon A \rightarrow
C$ a linear map. If for some $k \geq 2$, $\tau \otimes id_k \colon A
\otimes M_k(\bC) \rightarrow C \otimes M_k(\bC)$ is decomposable,
 then $\tau$ is completely positive.
\end{thm}

\vskip 2mm

We need one more lemma.

\begin{lem}\label{lem:completely copositive}
Let $A$ and $C$ be unital C*-algebras and let $\tau \colon A
\rightarrow C$ be a non-zero positive linear map. Then for $k\geq
2$, $\tau \otimes id_k \colon A \otimes M_k(\bC) \rightarrow C
\otimes M_k(\bC)$ is not completely copositive.
\end{lem}

\begin{proof}
Let $\{e_{ij}\}_{ij=1}^k$ be matrix units for $M_k(\bC)$. Suppose
that $\tau \otimes id_k$ is completely copositive. Then, since $1_A
\otimes (\sum_{ij=1}^k e_{ij} \otimes e_{ij}) = [1_A \otimes e_{ij}]
(=[b_{ij}] ) \in (A \otimes M_k(\bC))\otimes M_k(\bC)$ is positive
and $\tau \otimes id_k$ is completely copositive, $[\tau \otimes
id_k(b_{ji})]$ is positive by \cite{Stormer 1982}, that is,
$$
[\tau(1_A)  \otimes e_{ji}] = \tau(1_A) \otimes (\sum_{ij=1}^n e_{ji} \otimes e_{ij})
$$
is positive. But, $\sum_{ij=1}^n e_{ji} \otimes e_{ij}$ is not positive, and we have a contradiction.
\end{proof}

\vskip 2mm

\begin{rem}
The point of the previous lemma is that $id_k$ is not completely copositive. Hence, we have the following
observation: Let $A$ and $C$ be a unital C*-algebras and $\tau \colon A \rightarrow C$ be
 a non-zero positive linear map, and $\phi$ be not completely copositive  on $M_k(\bC)$.
 Then $\tau \otimes \phi \colon A \otimes M_k(\bC) \rightarrow C \otimes M_k(\bC)$ is not completely copositive.
\end{rem}

\vskip 2mm

{\it Proof of Theorem~\ref{thm:main2}}:

Suppose that $\tau \otimes id_k$ is decomposable. Then $\tau \otimes
id_k = \alpha + \beta$, for some $\alpha, \beta $, where $\alpha$ is
completely positive and $\beta$ is completely copositive. In
particular, $\alpha , \beta $ are positive. Then by
Theorem~\ref{factorization1} there exist positive linear maps
$\alpha_1 \colon A \rightarrow C$ and $\beta_1\colon A \rightarrow
C$ such that $\alpha = \alpha_1 \otimes id_k$ and $\beta = \beta_1
\otimes id_k$. From Lemma~\ref{lem:completely copositive} we know
that $\beta$ is not completely copositive. Hence, $\beta$ should be
zero. Hence $\tau = \alpha_1 $ and we have the conclusion.
\hfill$\qed$

\vskip 2mm
\section{Conclusions and Outlook}


We studied maps dominated by tensor products of  positive (or Completely Positive) maps.
Our main results are  Theorem
~\ref{factorization1} and ~\ref{factorization2}, where we show in two different contexts positive (or CP) maps dominated
by maps of the form $\alpha _1\otimes \alpha _2$, where $\alpha _2$ is pure, are necessarily of the form $\gamma \otimes \alpha _2$ for some $\gamma .$  An application of this to indecomposable maps is seen in the next Section. A general question about factorization of positive linear maps (Question ~ \ref{qst:split}) remains open.

In \cite{Terhal 2001} Terhal asked  as to whether there exist
$k$-positive  indecomposable maps from $M_m(\bC)$ to
 $M_n(\bC)$ for $1 < k < m \leq n$. As observed by Huber et al. \cite{HLLM}, a  partial answer to this question can be given using Theorem ~
\ref{thm:main2}, as follows. Suppose $\phi : M_a(\bC )\to M_b(\bC )$ is 4-positive but not CP (this requires $a,b\geq 5).$ Then
$\phi \otimes id_2: M_a(\bC)\otimes M_2(\bC) \to M_b(\bC )\otimes M_2(\bC) $
is  2-positive but not CP. Then by Theorem ~\ref{thm:main2}, $\phi \otimes id_2$ is also not decomposable. So we get an example of
an indecomposable map with $m,n\geq 10.$  However, any 2-positive map from $M_3(\bC)$ to $M_3(\bC)$ is decomposable by \cite{YLT 2018}.  The following question remains: Is there a 2-positive map from $M_m(\bC)$ to $M_n(\bC)$ for $4 \leq
m \leq n \leq 9$, which is not decomposable?

\vskip 2mm



\end{document}